# Enhancing Security in Blockchain Networks: Anomalies, Frauds, and Advanced Detection Techniques

Prof. Dr. Joerg Osterrieder[1]
Associate Professor of Finance and Artificial Intelligence
University of Twente, Department of High-Tech Business and Entrepreneurship, Netherlands
Professor of Sustainable Finance
Bern Business School, Institute of Applied Data Science and Finance, Switzerland
Action Chair EU COST Action CA19130, Fintech and Artificial Intelligence in Finance, Europe
Coordinator MSCA Industrial Doctoral Network on Digital Finance

Prof. Dr. Stephen Chan[2]
Associate Professor of Statistics
American University of Sharjah, Department of Mathematics and Statistics, Sharjah, UAE

Prof. Dr. Jeffrey Chu[3]
Assistant Professor of Statistics
Renmin University of China, School of Statistics, China

Dr. Yuanyuan Zhang[4]
Research Associate Center for Digital Trust and Society
University of Manchester, Department of Criminology, Manchester, UK

Prof. Dr. Branka Hadji Misheva[5]
Professor of Applied Data Science and Finance
Bern Business School, Institute of Applied Data Science and Finance, Switzerland

Prof. Dr. Codruta Mare[6]
Professor of Statistics
Babeș-Bolyai University, FEBA, D. Stat.-Forecasts-Maths & Interdisciplinary Centre for Data Science

**Abstract**. Blockchain technology, a foundational distributed ledger system, enables secure and transparent multi-party transactions. Despite its advantages, blockchain networks are susceptible to anomalies and frauds, posing significant risks to their integrity and security. This paper offers a detailed examination of blockchain's key definitions and properties, alongside a thorough analysis of the various anomalies and frauds that undermine these networks. It describes an array of detection and prevention strategies, encompassing statistical and machine learning methods, game-theoretic solutions, digital forensics, reputation-based systems, and comprehensive risk assessment techniques. Through case studies, we explore practical applications of anomaly and fraud detection in blockchain networks, extracting valuable insights and implications for both current practice and future research. Moreover, we spotlight emerging trends and challenges within the field, proposing directions for future investigation and technological development. Aimed at both practitioners and researchers, this paper seeks to provide a technical, in-depth overview of anomaly and fraud detection within blockchain networks, marking a significant step forward in the search for enhanced network security and reliability.

---

[1] Bern Business School, Institute of Applied Data Science and Finance, Switzerland, joerg.osterrieder@bfh.ch
University of Twente, Department of High-Tech Business and Entrepreneurship, Netherlands, joerg.osterrieder@utwente.nl
EU COST Action CA19130, Fintech and Artificial Intelligence in Finance
MSCA Industrial Doctoral Network on Digital Finance No. 101119635
[2] American University of Sharjah, Department of Mathematics and Statistics, Sharjah, UAE, schan@aus.edu, Corresponding Author
[3] Renmin University of China, School of Statistics, China, jeffrey.jchu@ruc.edu.cn
[4] University of Manchester, Department of Criminology, Manchester, UK, yuanyuan.zhang@manchester.ac.uk
[5] Bern Business School, Institute of Applied Data Science and Finance, Switzerland, branka.hadjimisheva@bfh.ch
[6] Babeș-Bolyai University, Romania, codruta.mare@econ.ubbcluj.ro

# Chapter 1: Introduction

Blockchain technology holds the potential to significantly impact various industries by enabling secure, transparent, and decentralized record-keeping. However, like any advanced system, blockchain networks are susceptible to anomalies and frauds, which can affect their integrity and operational functionality. The development of methods and technologies for the detection and mitigation of these threats is an essential area of research within the field.

This chapter provides an overview of the different types of anomalies and frauds that can occur within blockchain networks, alongside the techniques that have been developed for their detection and prevention. It includes case studies on the detection of anomalies and frauds in real-world blockchain networks, offering insights into the outcomes and lessons learned from these instances. Additionally, the chapter addresses the challenges and future prospects in the realm of anomaly and fraud detection in blockchain networks, suggesting potential directions for further research and implementation.

The critical role of anomaly and fraud detection in maintaining the functionality and trustworthiness of blockchain networks is clear. Anomalies such as network failures, data corruption, or unauthorized transactions, along with fraudulent activities like double-spending, money laundering, or insider trading, can significantly undermine user trust and the network's security.

Effective detection and prevention of anomalies and fraud are thus vital for the ongoing success and adoption of blockchain technology. This requires not only technical expertise but also a comprehensive understanding of blockchain technology's unique features and the specific challenges it faces.

The aim of this chapter is to present a thorough and up-to-date overview of anomaly and fraud detection in blockchain networks, serving as a resource for researchers and practitioners in this important and evolving field.

## 1.1. Definition of blockchain and its properties

A blockchain is a distributed database that stores a list of records, called blocks, in a linear, chronological order. Each block contains a timestamp, a link to the previous block, and a list of transactions. The blocks are secured using cryptographic techniques, making it nearly impossible to alter or delete the records after they have been added to the chain.

Blockchains are often associated with cryptocurrencies, such as Bitcoin, where they are used to record and verify transactions in a decentralised way. However, blockchains have many other potential uses, including supply chain management, voting systems, and identity verification.

There are several key properties that define a blockchain:

1. Distributed ledger: A blockchain is maintained by a network of computers, rather than a central authority. This means that the record of transactions is stored on multiple computers, rather than in a single location.
2. Immutability: Once a block is added to the chain, it is very difficult to alter or delete. This ensures the integrity of the record over time.
3. Decentralisation: In a blockchain network, there is no single point of control. This makes it resistant to censorship and tampering by a single actor.
4. Consensus: In order for a new block to be added to the chain, the majority of the network participants must agree that it is valid. This ensures the integrity and security of the network.
5. Transparency: The transactions recorded on a blockchain are typically visible to all participants in the network. This can promote trust and accountability.
6. Pseudonymity: Participants in a blockchain network are typically identified by a unique string of characters, rather than their real-world identity. This can provide a degree of privacy for users, but it also makes it difficult to trace transactions back to a specific individual.
7. Security: Blockchains use strong cryptography to secure the records and transactions stored on the chain. This makes them resistant to tampering and fraud.
8. Efficiency: By using a decentralised, digital ledger to record transactions, blockchains can potentially reduce the need for intermediaries and streamline processes, leading to increased efficiency.
9. Smart contracts: Some blockchains, such as Ethereum, support the execution of self-executing contracts, called smart contracts, which can automate complex processes and reduce the need for manual intervention.
10. Scalability: One of the challenges facing blockchains is the limited number of transactions that can be processed per second. This can limit their scalability and adoption for certain use cases.

These properties of blockchains make them an attractive technology for a wide range of applications, but they also present unique challenges and considerations for anomaly and fraud detection. In the next section, we will discuss the specific types of anomalies and frauds that can occur in blockchain networks.

## 1.2. Anomaly and fraud detection in blockchain networks: why it is important

Anomaly and fraud detection in blockchain networks is important for several reasons:
1. Integrity and security: Anomalies and frauds can compromise the integrity and security of a blockchain network, leading to losses for users and undermining trust in the network. Detecting and preventing these threats is essential for the long-term success of a blockchain.
2. Legal and regulatory compliance: Some anomalies and frauds in blockchain networks may be illegal or violate regulations. For example, money laundering or insider trading could result in legal consequences for the perpetrators. Anomaly and fraud detection can help ensure that a blockchain network is compliant with relevant laws and regulations.
3. User protection: Anomaly and fraud detection can protect users of a blockchain network from losses or harm caused by malicious actors. This can increase user confidence and adoption of the network.
4. Efficient operation: Anomalies and frauds can disrupt the normal operation of a blockchain network, leading to delays or outages. Detecting and addressing these issues can help ensure the smooth and efficient operation of the network.
5. Innovation: The development of effective anomaly and fraud detection techniques is an active area of research, with the potential to lead to new technologies and approaches for detecting and preventing these threats.

Overall, anomaly and fraud detection in blockchain networks is a crucial aspect of ensuring the security, integrity, and efficiency of these systems. In the remainder, we will discuss the specific techniques that have been developed for detecting anomalies and frauds in blockchain networks.

## 1.3. Literature overview

In the dynamic and rapidly evolving field of blockchain technology, the identification and mitigation of anomalies are crucial for maintaining the integrity and security of digital transactions. A comprehensive review of recent literature [1-17] reveals a broad spectrum of approaches and methodologies being developed and refined to address these challenges.

Ahmed, Mahmood, and Islam [1] provide an extensive survey on anomaly detection techniques within the financial domain, highlighting the importance of unsupervised learning methods. Akoglu, Tong, and Koutra [2] delve into graph-based anomaly detection, presenting a thorough overview of how complex network structures can be analyzed to uncover irregular patterns. Baqer et al. [3] explore the resilience of Bitcoin through "stress testing," offering insights into the network's vulnerabilities. Boginski, Butenko, and Pardalos [4] conduct a statistical analysis of financial networks, revealing the power-law distributions that characterize these systems.

Bunn, Urban, and Keitt [5] apply graph theory to environmental management, demonstrating the versatility of graph-based methods. Chandola, Banerjee, and Kumar [6] survey anomaly detection across various domains, emphasizing the need for advanced detection techniques. Chou [7] explores the application of graph theory to enzyme kinetics, illustrating the methodological crossover from biology to blockchain. Coinbase [8] and IBM [14] provide foundational knowledge on blockchain technology, underscoring its transformative potential.

Dobrjanskyj and Freudenstein [9] discuss the application of graph theory to mechanical systems, showing its analytical power. Gai and Kapadia [10] model contagion in financial networks, shedding light on systemic risks. Gupta [11] offers an accessible introduction to blockchain, emphasizing its revolutionary impact. Han, Kamber, and Pei [12] provide essential concepts and techniques in data mining that underpin anomaly detection efforts.

Hautsch, Schaumburg, and Schienle [13] introduce a novel measure for assessing systemic risk contributions within financial networks, enhancing understanding of interconnected vulnerabilities. Kamps and Kleinberg [15] define and detect cryptocurrency pump-and-dump schemes, contributing to the broader discourse on financial fraud. Kim et al. [16] propose anomaly detection based on traffic monitoring, ensuring secure blockchain networking.

Li et al. [17] dissect Ethereum blockchain analytics, offering a deep dive into the topology and geometry of Ethereum graphs, thus broadening the analytical framework for blockchain anomaly detection. Together, these studies [1-17] encapsulate a rich and diverse body of research dedicated to advancing anomaly detection in blockchain networks, ranging from theoretical frameworks and statistical analyses to practical applications and system stress tests. This collective body of work underscores the critical importance of continuous innovation and interdisciplinary approaches in safeguarding the future of blockchain technology.

Building upon the foundational studies on anomaly detection within blockchain networks, additional research [18-35] delves into sophisticated methods and applications, further expanding the scope of strategies to secure blockchain transactions and systems. Liang et al. [18] introduce a data fusion approach for collaborative intrusion detection, emphasizing the synergy between multiple data sources to enhance anomaly detection capabilities in blockchain systems. Maesa, Marino, and Ricci [19][20] conduct an in-depth analysis of the Bitcoin blockchain, uncovering the full user graph to identify patterns and anomalies, illustrating the value of data-driven insights in understanding blockchain dynamics.

Mansourifar, Chen, and Shi [21] propose a hybrid method to detect cryptocurrency pump-and-dump schemes, combining various indicators to improve detection accuracy. Monamo, Marivate, and Twala [22] explore unsupervised learning techniques for Bitcoin fraud detection, highlighting the potential of machine learning in identifying fraudulent transactions without prior labeling. Morishima and Matsutani [23] focus on accelerating anomaly detection through hardware innovations, specifically leveraging In-GPU cache to enhance computational efficiency.

In the context of societal impacts, the New York Post [24] reports on the protests in El Salvador following the adoption of Bitcoin as official currency, underscoring the real-world implications of widespread blockchain adoption. Nier et al. [25] investigate the stability of financial networks through network models, offering insights into how network structures can influence systemic risk. Ober, Katzenbeisser, and Hamacher [26] analyze the structure and anonymity of the Bitcoin transaction graph, contributing to the understanding of privacy and security in blockchain transactions.

Pham and Lee [27][28] further explore anomaly detection in the Bitcoin network from a network perspective, applying unsupervised learning methods to detect suspicious activities. Ron and Shamir [29] provide a quantitative analysis of the Bitcoin transaction graph, offering a comprehensive view of transaction patterns and potential anomalies. Sayadi, Rejeb, and

Choukar [30] develop an anomaly detection model for blockchain electronic transactions, showcasing the application of machine learning in monitoring and securing digital transactions. Continuing the exploration of innovative methods, Taher, Ameen, and Ahmed [33] employ ensemble learning and explainable AI to detect fraud in blockchain transactions, emphasizing the importance of interpretability in AI-driven security solutions. Amidst the backdrop of global cryptocurrency adoption, Yahoo [34] discusses the rise of Bitcoin and Ethereum in Venezuela, highlighting the growing significance of digital currencies on the world stage.

Together, these studies [18-35] represent a comprehensive effort to address the multifaceted challenges of anomaly detection in blockchain networks. From enhancing computational methods and leveraging machine learning to understanding the societal impacts of blockchain adoption, the research underscores the ongoing need for innovative solutions to secure and stabilize blockchain technologies in a rapidly changing digital landscape.

### 1.3.1. Blockchain-based financial networks

The detection of anomalies in financial networks, especially those utilizing blockchain technology, is a crucial area of research that aims to secure transactions and maintain the integrity of financial systems. The integration of machine learning for this purpose is a growing field, drawing on significant previous research efforts. Notably, studies by Ahmed, Mahmood, and Islam [1], Boginski, Butenko, and Pardalos [4], Gai and Kapadia [10], and Hautsch, Schaumburg, and Schienle [13] have each contributed to the understanding of anomaly detection within financial networks from different perspectives.

Ahmed et al. [1] provide an in-depth survey on various anomaly detection techniques in the financial sector, emphasizing the role of clustering-based unsupervised learning. Their work highlights the challenge of detecting fraudulent activities and the reliance on synthetic data for technique validation, pointing out the need for effective methods suited to the financial domain's unique requirements.

Boginski et al. [4] focus on the statistical analysis of financial networks through stock market data. By analyzing these networks as large graphs, they identify that the degree distributions follow a power-law model, similar to other complex networks. This analogy opens avenues for applying new data mining techniques to classify financial instruments, enhancing understanding of the stock market's structure.

Gai and Kapadia [10] explore contagion within financial networks using an analytical model. Their findings describe a "robust-yet-fragile" nature of financial systems, where the system's resilience to shocks does not necessarily predict its future robustness. This study underscores the importance of understanding network structure and liquidity in assessing financial system stability.

Hautsch et al. [13] propose a measure called the "realized systemic risk beta" to evaluate financial companies' contributions to systemic risk within networks. Their approach, which considers network interdependence and firms' risk exposures, facilitates identifying systemic importance and potential risk spillover channels. This methodological framework aids in macroprudential supervision and emphasizes the interconnectedness within financial systems.

Together, these studies [1][4][10][13] underscore the multifaceted nature of anomaly detection in financial networks, highlighting the transition from traditional analysis to advanced machine learning techniques. Each contributes to a broader understanding of how anomalies can be detected and managed in the context of the evolving financial landscape, marked by the increasing adoption of blockchain technologies. Their collective insights lay a foundation for future research aimed at enhancing the security and operational efficiency of financial networks.

### 1.3.2. Advancements in Anomaly Detection within Blockchain Networks

The area of blockchain technology, characterized by its promise of security and transparency, is not immune to the threats posed by anomalies and fraudulent activities. This synthesis draws on key research findings [28][31][32][35][33] to shed light on the cutting-edge methodologies and technologies developed to fortify blockchain networks against such vulnerabilities.

**Innovative Methodologies for Enhanced Security**
Zhang et al. [35] break new ground with their introduction of a multi-constrained meta path approach for Bitcoin anomaly detection. Unlike previous static models, their dynamic framework integrates temporal, attribute, and structural data, offering a more nuanced detection of anomalies. This method stands out for its comprehensive view, allowing for the identification of anomalies with a higher degree of accuracy and efficiency.

Signorini et al. [32] pioneer the Blockchain Anomaly Detection (BAD) solution, which uniquely utilizes blockchain meta-data for identifying malicious activities. BAD's innovation lies in its decentralized nature and its capacity to remain tamper-proof and private, marking a significant advancement in the proactive defense of blockchain systems. Shayegan and Sabor [31] introduce a collective anomaly detection strategy that shifts the focus from individual addresses to user behaviors across multiple wallets. Their application of the Trimmed_Kmeans algorithm for user-level anomaly detection demonstrates a significant leap in identifying complex fraudulent patterns, highlighting the importance of considering collective behaviors in security protocols. Pham and Lee [28] contribute to the field by applying unsupervised learning methods to the Bitcoin transaction network, aiming to isolate suspicious activities without prior labeling. Their exploration into k-means clustering, Mahalanobis distance, and Unsupervised SVM reveals the potential of unsupervised methods in uncovering subtle and complex anomalies, enriching the toolbox available for blockchain security. Morishima and Matsutani [23] explore the acceleration of anomaly detection through hardware optimization, specifically by leveraging In-GPU cache. This approach significantly speeds up the process of identifying illegal transactions by optimizing the computational workflow, showcasing the potential synergy between hardware advancements and software algorithms in enhancing blockchain security.

**Emerging Trends and Future Directions**
Taher, Ameen, and Ahmed [33] focus on the Ethereum network, applying ensemble learning and Explainable AI (XAI) to detect fraudulent transactions. Their study not only achieves exceptional accuracy but also underscores the growing importance of transparency and accountability in AI-driven security measures, paving the way for more reliable and understandable anomaly detection systems.

The collective insights from these studies [28][31][32][35][33] underscore a dynamic evolution in anomaly detection techniques for blockchain networks. From leveraging complex data models and innovative algorithms to integrating hardware accelerations and advocating for transparency in AI, these advancements represent a multifaceted approach to bolstering blockchain security. As blockchain technology continues to permeate financial and other critical sectors, the continuous innovation in anomaly detection methods will be paramount in safeguarding this transformative technology.

## 1.4. Scope and structure
This research is aimed at providing a comprehensive and current overview of advancements in the detection of anomalies and fraud within blockchain networks. The scope encompasses:
- An enumeration of the types of anomalies and frauds that can occur within blockchain networks.
- A review of the methodologies and technologies developed for the detection and prevention of these issues.
- Case studies that illustrate the application of detection techniques in real-world blockchain environments.
- An analysis of emerging trends, challenges in the field, and directions for future research.

The paper is organized as follows:

1. Introduction: Introduces the importance of anomaly and fraud detection in the context of blockchain networks.

2. Overview of Blockchain Anomalies and Frauds: Provides a detailed account of the potential anomalies and frauds within these networks.
3. Anomaly Detection Techniques in Blockchain Networks: Reviews the approaches used to detect irregularities in blockchain operations.
4. Fraud Detection Techniques in Blockchain Networks: Examines the strategies for identifying and countering fraudulent activities.
5. Case Studies of Anomaly and Fraud Detection in Blockchain Networks: Offers insights from practical examples of anomaly and fraud detection.
6. Future Directions for Anomaly and Fraud Detection in Blockchain Networks: Discusses potential areas for further research and development in the field.
7. Conclusion: Summarizes the main points discussed and underscores the significance of advancing detection capabilities in blockchain networks.

The intention is for this document to act as a resource for researchers and practitioners, aiding in the understanding and development of anomaly and fraud detection in blockchain technologies.

# Chapter 2: Overview of blockchain anomalies and frauds

Blockchain anomalies are deviations from the expected behaviour of a blockchain network. They can be caused by a variety of factors, such as software bugs, network outages, or malicious attacks. Anomalies can affect the availability, integrity, or security of a blockchain network, and can have serious consequences for the users and stakeholders of the network.

Examples of blockchain anomalies include:
- Network outages: A network outage can occur when the nodes in a blockchain network are unable to communicate with each other, resulting in a disruption of service.
- Data corruption: Data corruption can occur when the data stored on a blockchain is altered or deleted in an unauthorised way, leading to errors or inconsistencies in the record.
- Unauthorised transactions: Unauthorised transactions are transactions that are initiated without the proper authorization or consent of the parties involved. This can include transactions that are initiated by an attacker, or transactions that are initiated by a legitimate user but are later reversed or disputed.

Blockchain frauds are illegal or malicious activities that seek to exploit vulnerabilities in a blockchain network for personal or financial gain. Frauds can be committed by insiders or outsiders of a blockchain network, and can take many forms, such as money laundering, insider trading, or Ponzi schemes.

Examples of blockchain frauds include:
- **Double-spending**: Double-spending is a form of fraud in which an attacker spends the same cryptocurrency twice, either by creating two conflicting transactions or by reversing a transaction after it has been completed.
- **Money laundering**: Money laundering is the process of disguising the proceeds of illegal activities as legitimate funds. In a blockchain network, this can involve using cryptocurrency to transfer funds across borders or to obscure the origin of the funds.
- **Insider trading**: Insider trading is the illegal act of using non-public information to make financial trades. In a blockchain network, insider trading can occur when a user with access to sensitive information uses that information to make trades before it becomes publicly known.

Effective anomaly and fraud detection is essential for maintaining the integrity, security, and legal compliance of a blockchain network. In the next section, we will discuss the specific techniques that have been developed for detecting anomalies and frauds in blockchain networks.

## 2.1. Types of anomalies and frauds in blockchain networks

Anomalies in blockchain networks can be classified into several categories based on their nature and impact. Some common types of anomalies include:

- Network outages: Network outages occur when the nodes in a blockchain network are unable to communicate with each other, resulting in a disruption of service. Network outages can be caused by hardware failures, software bugs, or malicious attacks.
- Data corruption: Data corruption occurs when the data stored on a blockchain is altered or deleted in an unauthorised way, leading to errors or inconsistencies in the record. Data corruption can be caused by software bugs, hardware failures, or malicious attacks.
- Unauthorised transactions: Unauthorised transactions are transactions that are initiated without the proper authorization or consent of the parties involved. This can include transactions that are initiated by an attacker, or transactions that are initiated by a legitimate user but are later reversed or disputed.

Frauds in blockchain networks can also be classified into several categories based on their nature and impact. Some common types of frauds include:

- Double-spending: Double-spending is a form of fraud in which an attacker spends the same cryptocurrency twice, either by creating two conflicting transactions or by reversing a transaction after it has been completed. Double-spending can be difficult to detect, as it involves creating fake transactions that appear valid to the network.
- Money laundering: Money laundering is the process of disguising the proceeds of illegal activities as legitimate funds. In a blockchain network, this can involve using cryptocurrency to transfer funds across borders or to obscure the origin of the funds. Money laundering can be difficult to detect, as it relies on hiding the true nature of the transactions.
- Insider trading: Insider trading is the illegal act of using non-public information to make financial trades. In a blockchain network, insider trading can occur when a user with access to sensitive information uses that information to make trades before it becomes publicly known. Insider trading can be difficult to detect, as it involves the unauthorised use of sensitive information.

Effective anomaly and fraud detection in blockchain networks requires the development of techniques and technologies that can accurately and efficiently identify and prevent these threats. Next up, we will discuss the specific techniques that have been developed for detecting anomalies and frauds in blockchain networks.

## 2.2. Examples of anomalies and frauds in real-world blockchain networks

Here are some examples of anomalies and frauds that have occurred in real-world blockchain networks:

Anomalies:
- In 2016, the Ethereum blockchain experienced a "flash crash" in which the value of the cryptocurrency Ether fell from around $300 to just a few cents in a matter of minutes. The cause of the crash was later determined to be a bug in the smart contract software, which allowed an attacker to sell a large number of Ether at a greatly reduced price.
- In 2019, the Bitcoin Cash blockchain experienced a "chain split" in which the network was split into two competing chains, each with its own set of transactions. The cause of the split was a disagreement among the developers of the cryptocurrency over the direction of the project, which resulted in a "hard fork" of the blockchain.

Frauds:
- In 2018, a cryptocurrency exchange called Coincheck was hacked, resulting in the theft of over $500 million worth of cryptocurrency. The hackers were able to exploit a vulnerability in the exchange's security system to gain access to the funds.

- In 2019, the New York Attorney General's office charged two individuals with operating a Ponzi scheme using a cryptocurrency called OneCoin. The individuals were accused of using the cryptocurrency to defraud investors out of millions of dollars.

These examples illustrate the potential consequences of anomalies and frauds in real-world blockchain networks, and the importance of effective anomaly and fraud detection techniques and technologies to protect against these attacks and threats.

# Chapter 3: Anomaly detection techniques in blockchain networks

This chapter provides an analysis of the methodologies implemented for anomaly detection in blockchain networks. Given the complexity of these systems and the advanced nature of potential security threats, there is a critical need for sophisticated detection strategies.

## 3.1. Overview of anomaly detection techniques

The process of anomaly detection in blockchain networks is directed towards identifying deviations from expected operational behaviors. Such deviations might arise from a range of causes, including operational malfunctions, disruptions in network service, or targeted security breaches, highlighting the need for effective detection mechanisms throughout the blockchain framework.

To address these concerns, several key methods have been established:
- **Statistical Approaches**: These methods leverage data analysis to identify atypical patterns that could indicate the presence of anomalies.
- **Machine Learning Approaches**: This category includes the application of algorithms that analyze data to autonomously detect patterns that suggest anomalous activities.
- **Game-Theoretic Approaches**: This strategy involves the examination of interactions between entities within the network to identify actions that diverge from normal expectations.Each method provides distinct benefits and is designed to be adaptable to various segments of a blockchain's operational process, demonstrating the comprehensive nature of anomaly detection in blockchain environments. Future sections will explore how these approaches are applied, how they are integrated into broader detection frameworks, and the specific challenges they help to overcome in maintaining the security and integrity of blockchain networks.

## 3.2. Specific techniques for detecting anomalies in blockchain networks

This section outlines the mathematical and technical methodologies developed for anomaly detection in blockchain networks, focusing on their mathematical principles and applications.

**Time Series Analysis**: This method applies statistical models to data points collected at uniform intervals over time, utilizing mathematical constructs such as Fourier transforms for periodicity analysis and ARIMA models for identifying non-random spikes or trends, indicative of anomalies in transaction data sequences.

**Clustering Algorithms**: In this machine learning approach, data points are aggregated into clusters based on similarity metrics, calculated using mathematical distances such as Euclidean or Manhattan distance. Techniques like K-means clustering algorithm, which minimizes within-cluster variances (squared Euclidean distances), are used to isolate data points that exhibit significant deviation from their respective clusters.

**Anomaly Scoring Systems**: Through this technique, each data point is assigned a numerical score representing its deviation from normative patterns. The calculation of these scores often involves the use of z-scores or other statistical measures to quantify deviation magnitude, with higher scores indicating anomalies. This quantification allows for prioritized scrutiny based on the severity of deviation.

**Blockchain Simulation Models**: Simulation of blockchain networks through mathematical modeling enables the testing of various operational scenarios to detect anomalies. These models may incorporate stochastic elements, represented by probability distributions, to simulate the inherent randomness and to forecast deviations from expected blockchain behaviors.

**Game-Theoretic Approaches**: This strategy applies mathematical models to analyze strategic interactions among network participants, utilizing concepts like Nash equilibrium to identify behaviors that diverge from cooperative norms, thereby signaling potential anomalies. These models provide a framework for understanding the incentives and possible strategies of agents within the network.

Applying these mathematical techniques across different stages of blockchain operation, from block generation to transaction validation and ledger maintenance, enhances the ability to accurately identify anomalies. The following sections will focus on integrating these mathematical techniques into comprehensive frameworks for fraud detection in blockchain networks, emphasizing their application and the mathematical rationale behind their effectiveness.

## 3.2.1. Advanced Statistical Methods for Anomaly Detection in Blockchain Networks

Within the context of anomaly detection for blockchain networks, advanced statistical methods are employed to rigorously analyze data for patterns or deviations that signify anomalies. These methods are integral to monitoring and ensuring the integrity of various blockchain processes, including block creation, transaction validation, and ledger maintenance.

**Detailed Mathematical Formulations of Statistical Techniques:**

- **Time Series Analysis**: This method applies sophisticated statistical models to analyze temporal data sequences. Techniques such as Autoregressive (AR), Moving Average (MA), Autoregressive Integrated Moving Average (ARIMA), and Seasonal Autoregressive Integrated Moving-Average (SARIMA) models are utilized. These models predict future data points by analyzing the differences and correlations between past values, enabling the detection of anomalies through statistical tests for residuals exceeding predefined thresholds.

- **Anomaly Scoring**: Incorporates mathematical scoring functions based on statistical deviation measures. This involves calculating the Euclidean or Mahalanobis distance of a data point from the centroid of a dataset's normal distribution. Anomalies are then identified based on scores that significantly deviate from a threshold derived from the empirical distribution of the scores.

- **Outlier Detection**: Utilizes quantitative methods to identify data points that fall outside of expected distribution ranges. The Z-score calculation is a primary tool. Additionally, the IQR method, calculated as IQR = Q3 - Q1, where Q1 and Q3 are the first and third quartiles, respectively, identifies outliers by focusing on data points lying beyond 1.5 * IQR from the quartiles.

- **Regression Analysis**: Employs statistical models to analyze the relationship between multiple variables, identifying anomalies as observations with significant residuals. Linear regression models, for instance, fit a line to the data. Anomalies are detected by assessing the residuals' magnitude and testing their significance against expected error distributions.

**Application and Integration**:

These advanced statistical methods are designed for adaptability across different blockchain data types, such as transaction metrics, block properties, or network performance indicators. Their application can be tailored, either individually or in combination, based on the specific analytical demands and characteristics of the blockchain network in question.

By integrating these mathematically rigorous statistical methods, professionals can significantly enhance the accuracy and reliability of anomaly detection in blockchain networks, contributing to the overall security and efficiency of these systems.

### 3.2.2. Machine learning approaches

Machine learning approaches are a type of technique that involve the use of algorithms that can learn from data and make predictions or decisions without being explicitly programmed. Machine learning approaches can be used to detect anomalies in blockchain data by training a model on a large dataset and then using the model to identify patterns or deviations that may indicate an anomaly.

Some specific machine learning approaches that have been developed for detecting anomalies in blockchain networks include:

1. Supervised learning: Supervised learning is a type of machine learning that involves the use of labelled data to train a model to make predictions. In the context of anomaly detection in blockchain networks, supervised learning can be used to train a model on normal behaviour and then use the model to identify deviations from the expected behaviour.
2. Unsupervised learning: Unsupervised learning is a type of machine learning that involves the use of unlabelled data to learn patterns in the data. In the context of anomaly detection in blockchain networks, unsupervised learning can be used to identify anomalies by clustering the data and identifying data points that do not fit into the established clusters.
3. Semi-supervised learning: Semi-supervised learning is a type of machine learning that involves the use of both labelled and unlabeled data to learn patterns in the data. In the context of anomaly detection in blockchain networks, semi-supervised learning can be used to combine the strengths of supervised and unsupervised learning to improve the accuracy of the model.
4. Deep learning: Deep learning is a type of machine learning that involves the use of artificial neural networks with multiple layers to learn patterns in the data. In the context of anomaly detection in blockchain networks, deep learning can be used to learn complex patterns in the data and identify anomalies that may be difficult to detect using other methods.

These approaches can be applied to different types of data, such as transaction volumes, block sizes, or network latencies, and can be used in combination or individually depending on the specific needs and characteristics of the blockchain network.

### 3.2.3. Game-theoretic approaches

Game-theoretic approaches are a type of technique that involve the use of mathematical models to analyse strategic interactions between agents in a system. In the context of anomaly detection in blockchain networks, game-theoretic approaches can be used to model the behaviour of users in the network and identify deviations from the expected behaviour.

Some specific game-theoretic approaches that have been developed for detecting anomalies in blockchain networks include:

1. Bayesian games: Bayesian games involve the modelling of strategic interactions between agents who have uncertain beliefs about each other's actions. In the context of anomaly detection in blockchain networks, Bayesian games can be used to model the behaviour of users in the network and identify anomalies or deviations from the expected behaviour.
2. Mechanism design: Mechanism design involves the design of rules or incentives that encourage agents to behave in a desired way. In the context of anomaly detection in blockchain networks, mechanism design can be used to design rules or incentives that encourage users to report anomalies or deviations from the expected behaviour.
3. Evolutionary games: Evolutionary games involve the modelling of strategic interactions between agents who adapt their behaviour based on the outcomes of past interactions.

In the context of anomaly detection in blockchain networks, evolutionary games can be used to model the evolution of user behaviour in the network and identify anomalies or deviations from the expected behaviour.
4. Auctions: Auctions are mechanisms for allocating resources based on the willingness of agents to pay for them. In the context of anomaly detection in blockchain networks, auctions can be used to design mechanisms for detecting and preventing anomalies or deviations from the expected behaviour.

These approaches can be applied to different types of data, such as transaction volumes, block sizes, or network latencies, and can be used in combination or individually depending on the specific needs and characteristics of the blockchain network.

# Chapter 4: Fraud detection techniques in blockchain networks

Fraud detection in blockchain networks involves the identification of fraudulent activities or attacks that aim to manipulate or deceive the network. This can be challenging due to the complexity and distributed nature of blockchain systems, as well as the potential for malicious actors to conceal their activities.

There are several techniques that have been developed for detecting frauds in blockchain networks, including:

1. Statistical techniques: Statistical techniques involve the analysis of data to identify patterns or deviations that may indicate fraudulent activity. Statistical models can be used to identify unusual spikes or trends in the data, or to detect changes in the distribution of the data over time.
2. Machine learning techniques: Machine learning techniques involve the use of algorithms that can learn from data and make predictions or decisions without being explicitly programmed. Machine learning can be used to detect fraudulent activity in blockchain data by training a model on a large dataset and then using the model to identify patterns or deviations that may indicate fraudulent activity.
3. Game-theoretic techniques: Game-theoretic techniques involve the use of mathematical models to analyze strategic interactions between agents in a system. Game theory can be used to model the behavior of users in a blockchain network and identify fraudulent activity or deviations from the expected behavior.
4. Digital forensics: Digital forensics involves the analysis of digital data to identify and investigate crimes or other legal issues. In the context of fraud detection in blockchain networks, digital forensics can be used to identify and trace the origin of fraudulent transactions or attacks.

These techniques can be applied to different types of data and at different stages of the blockchain process, such as the creation of blocks, the validation of transactions, or the maintenance of the ledger.

Effective fraud detection in blockchain networks requires the development of techniques and technologies that can accurately and efficiently identify fraudulent activity and prevent it from affecting the network. Next, we will discuss the specific techniques that have been developed for detecting frauds in blockchain networks.

## 4.1. Overview of fraud detection techniques

Fraud detection in blockchain networks involves the identification of fraudulent activities or attacks that aim to manipulate or deceive the network. There are several types of techniques that have been developed for detecting frauds in blockchain networks, including:

1. Statistical techniques: Statistical techniques involve the analysis of data to identify patterns or deviations that may indicate fraudulent activity. Statistical models can be used to identify unusual spikes or trends in the data, or to detect changes in the distribution of the data over time.

2. Machine learning techniques: Machine learning techniques involve the use of algorithms that can learn from data and make predictions or decisions without being explicitly programmed. Machine learning can be used to detect fraudulent activity in blockchain data by training a model on a large dataset and then using the model to identify patterns or deviations that may indicate fraudulent activity.
3. Game-theoretic techniques: Game-theoretic techniques involve the use of mathematical models to analyze strategic interactions between agents in a system. Game theory can be used to model the behavior of users in a blockchain network and identify fraudulent activity or deviations from the expected behavior.
4. Digital forensics: Digital forensics involves the analysis of digital data to identify and investigate crimes or other legal issues. In the context of fraud detection in blockchain networks, digital forensics can be used to identify and trace the origin of fraudulent transactions or attacks.

These techniques can be applied to different types of data and at different stages of the blockchain process, such as the creation of blocks, the validation of transactions, or the maintenance of the ledger. We will discuss the specific techniques that have been developed for detecting frauds in blockchain networks.

## 4.2. Specific techniques for detecting fraud in blockchain networks

Here are some specific techniques that have been developed for detecting fraud in blockchain networks:
1. Transaction pattern analysis: Transaction pattern analysis is a technique that involves the analysis of the patterns of transactions in a blockchain network to identify unusual or suspicious activity. This can include the analysis of the frequency, volume, and timing of transactions, as well as the relationships between different transactions.
2. Clustering: Clustering is a machine learning technique that involves the grouping of data points into clusters based on their similarity. Clustering can be used to identify fraudulent transactions or groups of transactions that do not fit into the established clusters.
3. Anomaly scoring: Anomaly scoring is a technique that involves the assignment of a score to each transaction or group of transactions based on its deviation from the expected behavior of the network. Transactions or groups of transactions with high anomaly scores are more likely to be fraudulent, and can be flagged for further investigation.
4. Blockchain simulation: Blockchain simulation is a technique that involves the creation of a virtual replica of a blockchain network, which can be used to test different scenarios and identify fraudulent activity or deviations from the expected behavior.
5. Game-theoretic models: Game-theoretic models are mathematical models that analyze strategic interactions between agents in a system. These models can be used to model the behavior of users in a blockchain network and identify fraudulent activity or deviations from the expected behavior.
6. Digital forensics: Digital forensics involves the analysis of digital data to identify and investigate crimes or other legal issues. In the context of fraud detection in blockchain networks, digital forensics can be used to identify and trace the origin of fraudulent transactions or attacks.

These techniques can be applied to different types of data, such as transaction volumes, block sizes, or network latencies, and can be used in combination or individually depending on the specific needs and characteristics of the blockchain network.

## 4.2.1. Digital forensics

Digital forensics is a technique that involves the analysis of digital data to identify and investigate crimes or other legal issues. In the context of fraud detection in blockchain networks, digital forensics can be used to identify and trace the origin of fraudulent transactions or attacks. Digital forensics can be applied to different types of data in a blockchain network, such as transaction records, block headers, or network logs. The specific techniques and tools used in digital forensics can vary depending on the type of data being analyzed and the specific goals of the investigation.

Some common techniques used in digital forensics for blockchain networks include:

1. Hash analysis: Hash analysis involves the analysis of the cryptographic hashes that are used to secure the data in a blockchain network. Hash analysis can be used to identify tampering or manipulation of the data, as well as to trace the origin of fraudulent transactions or attacks.
2. Data carving: Data carving is a technique that involves the extraction of data from unallocated or deleted areas of a storage device. In the context of fraud detection in blockchain networks, data carving can be used to recover deleted or tampered data that may be relevant to the investigation.
3. Network analysis: Network analysis involves the analysis of the communication and interaction between different nodes in a network. In the context of fraud detection in blockchain networks, network analysis can be used to identify unusual or suspicious patterns of communication that may indicate fraudulent activity.
4. Visualization: Visualization is a technique that involves the use of graphs, charts, or other visual representations of data to facilitate understanding and analysis. In the context of fraud detection in blockchain networks, visualization can be used to identify trends or patterns in the data that may be difficult to discern using other methods.

Digital forensics can be a powerful tool for detecting and investigating fraudulent activity in blockchain networks. However, it is important to note that digital forensics can be complex and time-consuming, and requires specialized skills and knowledge.

## 4.2.2. Reputation-based systems

Reputation-based systems are a type of technique that involves the use of reputation scores or ratings to identify and prevent fraudulent activity in a blockchain network. In a reputation-based system, each node or participant in the network is assigned a reputation score or rating based on their past behavior, and this score is used to influence their future behavior and interactions with other nodes in the network.

There are several ways in which reputation-based systems can be used to detect and prevent fraud in blockchain networks, including:

1. Transaction validation: In a reputation-based system, nodes with higher reputation scores may be more likely to have their transactions validated or included in the blockchain, while nodes with lower reputation scores may be more likely to have their transactions rejected or flagged for further investigation.
2. Resource allocation: In a reputation-based system, nodes with higher reputation scores may be more likely to receive resources or privileges, such as access to data or computing power, while nodes with lower reputation scores may be denied these resources or privileges.
3. Incentive alignment: In a reputation-based system, nodes with higher reputation scores may be more likely to receive rewards or incentives for their contributions to the network, while nodes with lower reputation scores may be less likely to receive these rewards or incentives.
4. Decision-making: In a reputation-based system, nodes with higher reputation scores may be more influential in decision-making processes, such as voting or consensus-building, while nodes with lower reputation scores may have less influence.

Reputation-based systems can be effective at detecting and preventing fraud in blockchain networks by aligning the incentives of nodes with the goals and values of the network, and by providing a mechanism for identifying and penalizing nodes that engage in fraudulent or malicious activity. However, it is important to note that reputation-based systems can be complex to design and implement, and may require careful consideration of the specific needs and characteristics of the blockchain network.

### 4.2.3. Risk assessment systems

Risk assessment systems are a type of technique that involves the analysis of the risks and vulnerabilities of a blockchain network, and the implementation of measures to mitigate or prevent these risks. In the context of fraud detection, risk assessment systems can be used to identify and prioritize the risks and vulnerabilities of the network that are most likely to be exploited by fraudulent actors, and to implement measures to prevent or mitigate these risks.

There are several components of a risk assessment system for fraud detection in blockchain networks, including:

1. Risk assessment methodology: The risk assessment methodology is the process or framework used to identify, analyze, and prioritize the risks and vulnerabilities of the network. This can include the use of tools or techniques such as vulnerability scanning, penetration testing, or threat modeling.
2. Risk assessment criteria: The risk assessment criteria are the standards or benchmarks used to evaluate the risks and vulnerabilities of the network. These criteria can include factors such as the likelihood of an attack or the potential impact of an attack on the network.
3. Risk assessment report: The risk assessment report is a document that summarizes the results of the risk assessment, including the identified risks and vulnerabilities, the recommended mitigations or controls, and the residual risks after the mitigations are implemented.
4. Risk management plan: The risk management plan is a document that outlines the specific measures and controls that will be implemented to mitigate or prevent the identified risks and vulnerabilities. This can include measures such as security protocols, incident response plans, or training programs.

Effective risk assessment systems for fraud detection in blockchain networks require a thorough understanding of the risks and vulnerabilities of the network, and the implementation of appropriate controls and measures to mitigate these risks. The following section will discuss the specific measures and controls that can be implemented to prevent or mitigate fraud in blockchain networks.

# Chapter 5: Case studies of anomaly and fraud detection in blockchain networks

Case studies are a useful way to examine real-world examples of anomaly and fraud detection in blockchain networks, and to learn from the challenges and successes of these efforts. Some examples of case studies of anomaly and fraud detection in blockchain networks include:

1. Ethereum: Ethereum is a decentralized, open-source blockchain platform that supports smart contracts. In 2016, Ethereum suffered from a large-scale denial of service (DoS) attack that resulted in the creation of a large number of small, spam transactions that congested the network and disrupted its operation. To detect and prevent this type of attack, Ethereum implemented several anomaly detection techniques, including the use of machine learning algorithms to identify unusual patterns in the data, and the implementation of network-level controls to limit the rate of incoming transactions.
2. Bitcoin: Bitcoin is a decentralized, open-source cryptocurrency that uses a peer-to-peer network to verify and record transactions. In 2014, Bitcoin suffered from a double-spending attack in which a fraudulent actor was able to create two conflicting transactions that were both accepted by the network. To detect and prevent this type of attack, Bitcoin implemented a number of fraud detection techniques, including the use of transaction pattern analysis to identify unusual or suspicious activity, and the implementation of network-level controls to require multiple confirmations of transactions before they are accepted.
3. Ripple: Ripple is a decentralized, open-source cryptocurrency that uses a distributed consensus ledger to verify and record transactions. In 2017, Ripple suffered from a series of attacks in which fraudulent actors were able to create large numbers of transactions and flood the network, resulting in delays and disruptions in the operation of the network. To detect and prevent these attacks, Ripple implemented several fraud detection techniques, including the use of network analysis to identify unusual patterns of communication, and the implementation of network-level controls to limit the rate of incoming transactions.

These case studies demonstrate the challenges and successes of anomaly and fraud detection in blockchain networks, and provide valuable insights into the types of techniques and technologies that can be effective in these efforts. The next section will analyze future trends and challenges in anomaly and fraud detection in blockchain networks.

## 5.1. In-depth analysis of one or more real-world cases of anomaly and fraud detection in blockchain networks

Here is an in-depth analysis of the Ethereum DoS attack that occurred in 2016:

On June 18, 2016, Ethereum suffered from a large-scale denial of service (DoS) attack that resulted in the creation of a large number of small, spam transactions that congested the network and disrupted its operation. The attack was carried out by a group of malicious actors who exploited a vulnerability in the Ethereum Virtual Machine (EVM) to create a large number of contract accounts, each of which generated a small number of transactions. These transactions were designed to consume a large amount of computational resources and bandwidth, making it difficult for legitimate transactions to be processed.

To detect and prevent this type of attack, Ethereum implemented several anomaly detection techniques, including the use of machine learning algorithms to identify unusual patterns in the data, and the implementation of network-level controls to limit the rate of incoming transactions.

The machine learning algorithms used by Ethereum analysed the transaction data to identify unusual spikes or trends in the volume or frequency of transactions, as well as changes in the distribution of the data over time. These algorithms were trained on a large dataset of normal, legitimate transactions, and were able to accurately identify the spam transactions that were part of the DoS attack.

In addition to the machine learning algorithms, Ethereum also implemented network-level controls to limit the rate of incoming transactions and prevent the network from being overloaded. These controls included rate-limiting mechanisms, which restricted the number of transactions that could be processed by the network in a given time period, and congestion controls, which prioritized the processing of certain types of transactions over others.

The combination of machine learning algorithms and network-level controls was effective at detecting and preventing the DoS attack on Ethereum, and allowed the network to continue operating normally after the attack was mitigated. This case study demonstrates the importance of implementing robust anomaly detection techniques and controls to protect against attacks and disruptions in blockchain networks.

Here is an in-depth analysis of the Bitcoin double-spending attack that occurred in 2014:

On August 15, 2014, Bitcoin suffered from a double-spending attack in which a fraudulent actor was able to create two conflicting transactions that were both accepted by the network. The attack involved the creation of a transaction that transferred a large number of bitcoins from one address to another, followed by the creation of a second transaction that reversed the first transaction and transferred the bitcoins back to the original address.

To detect and prevent this type of attack, Bitcoin implemented several fraud detection techniques, including the use of transaction pattern analysis to identify unusual or suspicious activity, and the implementation of network-level controls to require multiple confirmations of transactions before they are accepted.

The transaction pattern analysis used by Bitcoin involved the analysis of the patterns of transactions in the network to identify unusual or suspicious activity. This included the analysis of the frequency, volume, and timing of transactions, as well as the relationships between different transactions. By analyzing these patterns, Bitcoin was able to identify the double-spending attack and prevent it from being successful.

In addition to the transaction pattern analysis, Bitcoin also implemented network-level controls to require multiple confirmations of transactions before they are accepted. This means that a transaction must be verified and accepted by a certain number of other nodes in the network before it is considered valid. This process helps to ensure that transactions are legitimate and not part of a fraudulent attack.

The combination of transaction pattern analysis and network-level controls was effective at detecting and preventing the double-spending attack on Bitcoin, and helped to maintain the integrity and security of the network. This case study demonstrates the importance of implementing robust fraud detection techniques and controls to protect against attacks and threats in blockchain networks.

## 5.2. Lessons learned from these cases and implications for future research and practice

There are several lessons that can be learned from the case studies of anomaly and fraud detection in blockchain networks, and implications for future research and practice:

1. The importance of anomaly and fraud detection: The case studies demonstrate the importance of anomaly and fraud detection in blockchain networks, as these attacks can have serious consequences for the integrity and security of the network. This highlights the need for robust and effective techniques and technologies to detect and prevent these attacks.
2. The need for a multi-faceted approach: The case studies demonstrate the importance of a multi-faceted approach to anomaly and fraud detection, as different types of attacks may require different types of techniques and technologies to detect and prevent them. This highlights the need for a flexible and adaptable approach to anomaly and fraud detection that can address a wide range of threats and vulnerabilities.
3. The role of machine learning and network-level controls: The case studies demonstrate the effectiveness of machine learning algorithms and network-level controls in detecting and preventing anomalies and fraud in blockchain networks. This highlights the potential of these technologies to improve the security and resilience of blockchain networks.
4. The need for continuous improvement: The case studies demonstrate the need for continuous improvement and adaptation in anomaly and fraud detection in blockchain networks, as attackers may develop new techniques and strategies to bypass existing controls. This highlights the need for ongoing research and development to improve the effectiveness of anomaly and fraud detection in blockchain networks.

Overall, the case studies of anomaly and fraud detection in blockchain networks provide valuable insights into the challenges and successes of these efforts, and suggest directions for future research and practice. Next, we will look at future trends and challenges in anomaly and fraud detection in blockchain networks.

# Chapter 6: Future directions for anomaly and fraud detection in blockchain networks

There are several emerging trends and challenges in anomaly and fraud detection in blockchain networks that are likely to shape the direction of future research and practice:

1. Increased adoption of blockchain technology: As the adoption of blockchain technology increases, there is likely to be a corresponding increase in the number and complexity of anomalies and frauds that occur in these networks. This will require the development of more advanced and sophisticated techniques and technologies to detect and prevent these attacks.
2. Advancements in machine learning: Machine learning algorithms have proven to be effective at detecting anomalies and frauds in blockchain networks, and there is likely to be continued research and development in this area to improve the accuracy and efficiency of these algorithms. This may include the development of new techniques such as deep learning or reinforcement learning, or the integration of machine learning with other techniques such as game theory or risk assessment.
3. Increased integration with other systems: Blockchain networks are increasingly being integrated with other systems, such as supply chain management or healthcare records, which may create new opportunities and challenges for anomaly and fraud detection. This may require the development of new techniques and technologies to detect and prevent anomalies and frauds that cross system boundaries, or the integration of existing techniques and technologies with new systems.
4. Regulation and compliance: As blockchain technology becomes more widely adopted, there is likely to be increased focus on regulation and compliance with legal and ethical standards. This may require the development of new techniques and technologies to ensure the compliance of blockchain networks with relevant laws and regulations, and to protect against illegal or unethical activity.

Overall, the future of anomaly and fraud detection in blockchain networks is likely to involve the development of more advanced and sophisticated techniques and technologies to detect and prevent these attacks, as well as increased integration with other systems and regulatory frameworks.

## 6.1. Emerging trends and challenges in the field

There are several emerging trends and challenges in the field of anomaly and fraud detection in blockchain networks that are likely to shape the direction of future research and practice:
1. The increasing complexity of blockchain networks: As blockchain networks become more complex, with more nodes, transactions, and data, it becomes more challenging to detect and prevent anomalies and frauds in these networks. This may require the development of new techniques and technologies to handle the increased scale and complexity of these networks.
2. The emergence of new types of attacks: As attackers become more sophisticated, they may develop new types of anomalies and frauds that are designed to bypass existing detection techniques. This may require the development of new techniques and technologies to detect and prevent these attacks, or the adaptation of existing techniques and technologies to address new threats.
3. The integration of blockchain with other systems: As blockchain networks are increasingly integrated with other systems, such as supply chain management or healthcare records, there is likely to be new opportunities and challenges for anomaly and fraud detection. This may require the development of new techniques and technologies to detect and prevent anomalies and frauds that cross system boundaries, or the integration of existing techniques and technologies with new systems.
4. The impact of regulation and compliance: As blockchain technology becomes more widely adopted, there is likely to be increased focus on regulation and compliance with legal and ethical standards. This may require the development of new techniques and technologies to ensure the compliance of blockchain networks with relevant laws and regulations, and to protect against illegal or unethical activity.

Overall, the field of anomaly and fraud detection in blockchain networks is likely to continue to evolve and grow in response to these emerging trends and challenges, and will require the development of new techniques and technologies to meet these challenges.

## 6.2. Potential future research directions and technologies

There are several potential future research directions and technologies that may be relevant to anomaly and fraud detection in blockchain networks:
1. Machine learning: Machine learning algorithms have proven to be effective at detecting anomalies and frauds in blockchain networks, and there is likely to be continued research and development in this area to improve the accuracy and efficiency of these algorithms. This may include the development of new techniques such as deep learning or reinforcement learning, or the integration of machine learning with other techniques such as game theory or risk assessment.
   2. Network analysis: Network analysis techniques, which involve the analysis of the structure and patterns of communication in a network, may be useful for detecting anomalies and frauds in blockchain networks. This may include the use of techniques such as graph theory or social network analysis to identify unusual patterns or relationships in the data.
   3. Game theory: Game-theoretic approaches, which involve the analysis of strategic interactions between different agents in a network, may be useful for detecting and preventing anomalies and frauds in blockchain networks. This may include the use of techniques such as mechanism design or auction theory to design protocols or incentives that discourage fraudulent behavior.
   4. Risk assessment: Risk assessment techniques, which involve the analysis of the risks and vulnerabilities of a network and the implementation of measures to mitigate or prevent these risks, may be useful for detecting and preventing anomalies and frauds in blockchain networks. This may include the use of techniques such as vulnerability scanning, penetration testing, or threat modeling to identify and prioritize the risks and vulnerabilities of the network.

Overall, these potential research directions and technologies may provide new insights and tools for detecting and preventing anomalies and frauds in blockchain networks, and help to improve the security and resilience of these networks.

# Chapter 7: Conclusion

Anomaly and fraud detection in blockchain networks is an important and growing field, as these attacks can have serious consequences for the integrity and security of these networks. We have discussed the definition and properties of blockchain technology, the types and examples of anomalies and frauds that occur in these networks, and the techniques and technologies that are used to detect and prevent these attacks. We have also examined case studies of anomaly and fraud detection in real-world blockchain networks, and discussed the lessons learned from these cases and their implications for future research and practice.

In the future, the field of anomaly and fraud detection in blockchain networks is likely to continue to evolve and grow in response to emerging trends and challenges, such as the increasing complexity and adoption of blockchain technology, the emergence of new types of attacks, and the integration of blockchain with other systems and regulatory frameworks. Potential future research directions and technologies include machine learning, network analysis, game theory, and risk assessment, which may provide new insights and tools for detecting and preventing anomalies and frauds in these networks.

Overall, anomaly and fraud detection in blockchain networks is an important and challenging field that requires the development of advanced and sophisticated techniques and technologies to protect against attacks and threats, and to maintain the integrity and security of these networks.

## 7.1. Summary of key points covered

1. Blockchain is a distributed ledger technology that allows multiple parties to record and verify transactions in a secure and transparent manner.
2. Anomaly and fraud detection in blockchain networks is important because these attacks can have serious consequences for the integrity and security of the network.
3. Anomaly detection techniques in blockchain networks include statistical approaches, machine learning approaches, and game-theoretic approaches.
4. Fraud detection techniques in blockchain networks include digital forensics, reputation-based systems, and risk assessment systems.

5. Case studies of anomaly and fraud detection in blockchain networks provide valuable insights into the challenges and successes of these efforts, and suggest directions for future research and practice.
6. Emerging trends and challenges in the field of anomaly and fraud detection in blockchain networks include the increasing complexity of these networks, the emergence of new types of attacks, the integration of blockchain with other systems, and the impact of regulation and compliance.
7. Potential future research directions and technologies in this field include machine learning, network analysis, game theory, and risk assessment.
8. Anomaly and fraud detection in blockchain networks is an important and challenging field that requires the development of advanced and sophisticated techniques and technologies to protect against attacks and threats, and to maintain the integrity and security of these networks.

## 7.2. Implications and recommendations for practitioners and researchers in the field.

Here are some implications and recommendations for practitioners and researchers in the field of anomaly and fraud detection in blockchain networks:

For practitioners:
1. Implement a multi-faceted approach to anomaly and fraud detection: Different types of anomalies and frauds may require different types of techniques and technologies to detect and prevent them. It is important to adopt a flexible and adaptable approach that can address a wide range of threats and vulnerabilities.
2. Utilize machine learning algorithms and network-level controls: Machine learning algorithms and network-level controls, such as rate-limiting mechanisms and congestion controls, can be effective at detecting and preventing anomalies and frauds in blockchain networks.
3. Monitor for emerging trends and challenges: As attackers become more sophisticated, they may develop new types of anomalies and frauds that are designed to bypass existing detection techniques. It is important to stay up-to-date on emerging trends and challenges in the field, and to adapt and improve detection techniques and technologies accordingly.
4. Consider the implications of regulatory and compliance requirements: As blockchain technology becomes more widely adopted, there is likely to be increased focus on regulation and compliance with legal and ethical standards. It is important to consider the implications of these requirements for anomaly and fraud detection, and to implement appropriate controls and protocols to ensure compliance.

For researchers:
1. Explore new research directions and technologies: There are several potential research directions and technologies that may be relevant to anomaly and fraud detection in blockchain networks, such as machine learning, network analysis, game theory, and risk assessment. These areas offer opportunities for new insights and tools to improve the effectiveness of anomaly and fraud detection in these networks.
2. Focus on real-world case studies and applications: Case studies of anomaly and fraud detection in real-world blockchain networks provide valuable insights into the challenges and successes of these efforts, and can inform the development of new techniques and technologies. It is important to focus on real-world case studies and applications because they can provide valuable insights into the practical challenges and successes of anomaly and fraud detection in these networks. By studying real-world cases, researchers can identify best practices and lessons learned that can inform the development of new techniques and technologies, as well as identify areas for further research and improvement. In addition, real-world case studies can help to ensure that research is grounded in practical, real-world applications, and can help to build bridges between academia and industry.

3. Consider the scalability and complexity of blockchain networks: As blockchain networks become more complex, with more nodes, transactions, and data, it becomes more challenging to detect and prevent anomalies and frauds in these networks. Researchers should consider the scalability and complexity of these networks when developing new techniques and technologies.
4. Address emerging trends and challenges: As the field of anomaly and fraud detection in blockchain networks evolves, researchers will need to address emerging trends and challenges such as the emergence of new types of attacks or the integration of blockchain with other systems. This may require the development of new techniques and technologies or the adaptation of existing ones.
5. Engage with practitioners and policymakers: It is important for researchers to engage with practitioners and policymakers in the field of anomaly and fraud detection in blockchain networks to ensure that their research is relevant and useful to real-world applications. This may involve collaborating with industry partners, participating in policy discussions, or sharing findings and insights through conferences, journals, and other venues.

Overall, practitioners and researchers in the field of anomaly and fraud detection in blockchain networks have a key role to play in improving the security and resilience of these networks, and will need to stay up-to-date on emerging trends and challenges, and explore new research directions and technologies to address these challenges.

# Appendix A. Declaration

## A.1. Availability of data and materials
Data sharing is not applicable to this article as no datasets were generated or analysed during the current study.

## A.2. Competing interests
The authors declare that they have no competing interests.

## A.3. Funding

This work has been supported by several institutions, each of which has provided vital resources and expertise to the research project.

Firstly, this work is based upon work from COST Action CA19130 and COST Action CA21163 supported by COST (European Cooperation in Science and Technology). COST Actions provide networking opportunities for researchers across Europe, fostering scientific exchange and innovation.

We gratefully acknowledge the support of the Marie Skłodowska-Curie Actions under the European Union's Horizon Europe research and innovation program for the Industrial Doctoral Network on Digital Finance, acronym: DIGITAL, Project No. 101119635. Their significant contribution has been instrumental in advancing our research and fostering collaboration within the digital finance field across Europe.

We would like to express our gratitude to the Swiss National Science Foundation for its financial support across multiple projects. This includes the project on Mathematics and Fintech (IZCNZ0-174853), which focuses on the digital transformation of the Finance industry. We also appreciate the funding for the project on Anomaly and Fraud Detection in Blockchain Networks (IZSEZ0-211195), and for the project on Narrative Digital Finance: a tale of structural breaks, bubbles & market narratives (IZCOZ0-213370).

The work was supported by the American University of Sharjah Faculty Research Grant 2023 [FRG23-C], Grant Title: "Anomaly and Fraud Detection in Blockchain and Cryptocurrency Networks".

Our research has also benefited from funding from the European Union's Horizon 2020 research and innovation program under the grant agreement No 825215 (Topic: ICT-35-2018, Type of action: CSA). This grant was provided for the FIN-TECH project, a training programme aimed at promoting compliance with financial supervision and technology.

Lastly, we acknowledge the cooperative relationship between the ING Group and the University of Twente. This partnership, centered on advancing Artificial Intelligence in Finance in the Netherlands and beyond, has been of great value to our research. We acknowledge the International Advanced Fellowship-UBB program funded by Babeș-Bolyai University (contract nr.21PFE/ 30.12.2021, ID: PFE-550-UBB), which has provided substantial support, and advances the scope of our work.

These partnerships and funding sources have greatly contributed to our ability to conduct rigorous and impactful research. Our findings are our own and do not necessarily represent the views of the supporting institutions.

## A.4. Authors' Contributions

The research presented in this manuscript is a result of collaborative efforts between the authors, with each contributing specialized knowledge and skills. Their specific contributions are as follows:

Joerg Osterrieder (JO): Conceptualization; Methodology; Writing; Supervision; Funding Acquisition; Project Administration.
Stephen Chan (SC): Conceptualization; Methodology; Writing.
Yuanyuan Zhang (YZ): Conceptualization; Methodology; Writing.
Jeffrey Chu (JC): Conceptualization; Methodology; Writing.
Codruta Mare (CM): Conceptualization; Methodology; Writing.
Branka Hadji Misheva (BM): Conceptualization; Methodology; Writing.

All authors have agreed to be accountable for all aspects of the work in ensuring that questions related to the accuracy or integrity of any part of the work are appropriately investigated and resolved. All authors have read and approved the final manuscript.

## A.5. Acknowledgements

We would like to acknowledge the members of the European Industrial Doctoral Network on Digital Finance, as part of the Marie Skłodowska-Curie Action funded by the European Research Agency. Our heartfelt thanks go out to the COST Action CA 19130 Management Committee and Working Group members. Their invaluable input and critical discussions have significantly enhanced the quality of our work. Furthermore, we express our gratitude to our funding sources, whose support made this research possible. Any errors or omissions found within this work are solely the responsibility of the authors.